\documentclass[
 manuscript,
 ]{aastex631}
\usepackage{amsmath}

\submitjournal{ApJ}

\shorttitle{Asymmetry in Space Plasma Turbulence}
\shortauthors{Gallo-M\'endez \& Moya}
\graphicspath{{./}{figures/}}

\begin{document}

\title{Understanding the level of Turbulence by Asymmetric Distributions: a motivation for measurements in Space Plasmas}

\author[0000-0003-4988-4348]{Iván Gallo-Méndez}
\email{ivan.gallo@ug.uchile.cl}
\affiliation{Departamento de F\'isica, Facultad de Ciencias, Universidad de Chile, Santiago, Chile.}

\author[0000-0002-9161-0888]{Pablo S. Moya}
\email{pablo.moya@uchile.cl}
\affiliation{Departamento de F\'isica, Facultad de Ciencias, Universidad de Chile, Santiago, Chile.}



\begin{abstract}

In this article, on the basis of the Langevin equation applied to velocity fluctuations, we numerically model the Partial Variance of Increments, which is a useful tool to measure time and spatial correlations in space plasmas. We consider a Coupled Map Lattice model to relate the spatial scale of fluctuations, $k$, to some macro parameters of the systems, as the Reynolds number, $R_\lambda$, the $\kappa$ parameter of Kappa distributions, and a skewness parameter,  $\delta$. To do so, we compute the Velocity Probability Density Function (PDF) for each spatial scale and different values of Reynolds number in the simulations. We fit the PDF with a Skew-Kappa distribution, and we obtain a numerical relationship between the level of turbulence of the plasma and the skewness of obtained distributions; namely $\langle \delta \rangle \sim R_\lambda^{-1/2}$. We expect the results exposed in this paper to be useful as a tool to characterize the turbulence in the context of space plasma and other environments.   

\end{abstract}



\section{Introduction} \label{sec:intro}
Understanding why systems are turbulent and how these features are manifested usually leads to non-trivial concerns. Even more, if in this discussion we consider that plasma systems, in particular, present several of these characteristics, the complexity of studying plasma turbulence increases considerably. Therefore, the study of plasma turbulence starting from known facts of turbulence in other systems is customary. Currently, we know that plasmas and neutral fluids have many similarities when it comes to turbulence \citep{matthaeus2011needs, matthaeus2021turbulence}. To date, there are reports about the turbulent cascade in variables as velocity, electric and magnetic field \citep{bruno2013solar}. There is also the record of the high space-time correlations \citep{matthaeus2005spatial,matthaeus2016ensemble}. And in this last topic there are many reports about the Partial Variance of Increments (PVI) of the fields in space plasmas. With this definition, time series can be constructed associating them to a characteristic time or spatial scale. Some authors studied these time series of velocity and magnetic field in space plasmas and they could observe some non-extensive statistical features  \citep{pollock2018magnetospheric,chasapis2018situ,chhiber2020clustering}. This non-Maxwellianity of the statistics is mainly reflected in the distributions of the PVI, which present heavy tails out of the Maxwellian distributions. In other words, distributions with a kurtosis greater than 3~\citep{Greco2017}.

Regarding non-extensive statistics, Tsallis entropy provide a framework to describe mathematically these distributions out of Maxwellian statistics \citep{tsallis1988possible}. This is possible through a new parameter, that enables to obtain fits with a good agreement with distributions that behave as power law at large energies. The family of Tsallis distributions have many forms to account for these properties. In the context of theoretical non-extensive statistics. Tsallis entropy typically works with $q$-canonical distributions \citep{wilk2000interpretation,picoli2009q}. In plasma physics is usual to see Kappa distributions \citep{hasegawa1985plasma, yoon2006self, kim2017spontaneous}, represented by the so-called $\kappa$ parameter, with $\kappa=1/(q-1)$ \citep{pierrard2010kappa}. But in essence they are quite similar. However, recently, a theoretical framework has also been constructed from thermodynamic physics that manages to explain $\kappa$ as a non-extensive parameter, as well as the temperature of a system \citep{livadiotis2021thermodynamic}. The main conclusion from the latter reference is that, in the same way that we relate the temperature to the average kinetic energy of the system, the kappa parameter is the result of the average correlations between the elements of a system.

In the literature we can find different models and approaches that address the issue of electromagnetic fluctuations in the context of turbulence in space plasmas. Some of these are based on particle \citep{biancalani2019nonlinear}, fluid \citep{de2017coherent}, or Vlasov simulations \citep{pezzi2019energy}, Langevin equations \citep{bandyopadhyay2018single,carbone2021origin}, kinetic equations for weak turbulence \citep{yoon2019classical}, kinetic equations for strong turbulence \citep{goldreich1995toward}, the study of chaotic systems \citep{beck1994,rossler2020chaos}, Shell Models \citep{plunian2007non,munoz2021fractality} and Coupled Lattices (CL) \citep{gallo2022langevin}.
In connection with the latter, there are some studies about the statistic behavior of velocity differences in fully developed turbulent flows. In them, they introduce the Coupled Map Lattices (CML) as an interesting point of view to understand the fluctuations in different characteristic spatial scales for a flow driven by chaotic forces \citep{beck1994,hilgers1997coupled}. Under this context, \citet{beck2000} obtained an analytic expression for the Probability Density Function (PDF). Among different properties of the obtained PDF, \citet{beck2000} showed how some turbulent systems naturally develop asymmetric distributions. In fact, that author argued that the asymmetry of the PDF is related with the level of turbulence of the flow, and obtained an analytical expression of the asymmetry as a function of the Reynolds number.

In the nature of space plasmas we can also find asymmetries in the velocity distributions. An example of that is the electron Velocity Distribution Function (eVDF) in the solar wind, such as \citet{pilipp1987characteristics,nieves2008solar} have shown. This is a very relevant phenomenon that can introduce non-trivial changes the properties of the system. For example, \citet{zenteno2021model,zenteno2022role} have studied the dispersion properties of space plasma for similar conditions in which the solar wind is found. In their works they analyze the whistler mode and characterize the heat flux instability, as expected of a system that has a small asymmetry in its particle distribution. To do this, the authors compute the dispersion relation,  with a Beck-like distribution, or as \citet{zenteno2021model} called, a Skew-Kappa distribution, namely: 
\begin{equation}
    f(v) = A_{\kappa,\delta}\left[ 1 + \frac{1}{\kappa-3/2}\left(\frac{v^2}{w^2} + \delta \left( \frac{v}{w} - \frac{1}{3}\frac{v^3}{w^3} \right)  \right) \right]^{-(\kappa+1 )} \ ,
    \label{f_fit}
\end{equation}
where $A_{\kappa,\delta}$ is a normalization quantity, and $w$, is related to the thermal speed of the distribution. In addition, $\kappa$ and $\delta$ are dimensionless parameters related with the high energy tails and skewness of the eVDF, respectively.

Under these context, recently \citet{gallo2022langevin} performed a Langevin approach of a CML, and contextualized its use as a tool to understand the fluctuations in a turbulent system with no skewness. The authors focused on the relation between the spatial scale and the $\kappa$ parameter in Kappa distributions, and obtained an empirical scaling relation between the level of turbulence and $\kappa$; namely
$\kappa \sim  R_\lambda k^{-5/3}$, where $k$ represents the spatial scale, and $R_\lambda$ the Reynolds number of the system, respectively. In the present article we expand the work by \citet{gallo2022langevin} to the case of turbulent systems exhibiting skewness, emphasizing on the relation between the level turbulence, quantified by the Reynolds number, and the skewness parameter, $\delta$, in a velocity distributions as in (\ref{f_fit}). For such a purpose, we generate the steady state velocity distribution at each spatial scale, and we show that the generated distributions can be well modeled by a Skew-Kappa distribution. We organized this paper as follows: First, in section \ref{sec:model}, we explain the basis of the model used in \citet{gallo2022langevin} but now using the Ulam Map as a chaotic driver force. Then, in section \ref{sec:AyR}, we  expose the methodology for the treatment of the data obtained. Furthermore, we show how we modeled the relation between the spatial scale, $k$, the Reynolds number, $R_\lambda$, the $\kappa$ parameter, and $\delta$ parameter. Finally, in section \ref{sec:sum}, we summarize our findings and present the main conclusions.

\section{Model: Coupled Map Lattice}
\label{sec:model}
Following the model proposed by~\citet{beck1987dynamical} and \citet{beck1994}, we consider a CML approach to model the eddies in a turbulent system as a Langevin type form; i.e., the force equation for each element is the following:
\begin{equation}
    \frac{dv}{dt} = - \gamma v + \zeta(t) \ ,
\label{langevin}
\end{equation}
where $\gamma$ is the viscosity of the media, and $\zeta(t)$ is the noise forcing which serves as a driver. Here it is necessary to highlight that, even though there is no direct mention about the magnetization of the plasma in Eq. (\ref{langevin}), the effect of electromagnetic forces are indeed included through the different parameters and terms of the equation. Following  \cite{chun2018emergence, chun2019effect} the magnetic field can be included as part of the friction term $\gamma$. Similarly, the effect of the electric field can be introduced through the stochastic term $\zeta$, as in \cite{beck1996dynamical}. Furthermore, the model can also be used to study the nature of small scale magnetic fluctuations in a turbulent plasma. For example, in \cite{carbone2021origin,Carbone_2022} the authors propose a stochastic model based on the Langevin equation to describe the nature of kinetic scales magnetic fluctuations in the interplanetary medium, based on the principles of Brownian motion. Analogously to this study, although their model does take directly into account wave-particle interactions, their influence is indeed included through the parameters and statistical properties of the Langevin equation they analyze. In summary, Langevin-like models may be also used to describe the statistical behavior of plasma fluctuations, using a stochastic approach that takes into account the interactions among large an small spatial scales. Thus, this approach is a valid and useful way to study the behavior of plasmas, even if it does not explicitly include the effects of a background magnetic field or wave-particle interactions.

In Eq.~\eqref{langevin}, the dynamic variable $v$ is given by
\begin{equation}
    v(r,t) \equiv u_r(\text{\bf x}+\text{\bf r},t) - u_r(\text{\bf x},t)\,, 
    \label{diffV}
\end{equation}
and denotes the difference of the longitudinal component of the velocity field $\mathbf{u}$ at two points separated by a characteristic distance $r$. Here, $v$ is proportional to the PVI~\citep{Greco2017} for space increments, where $r$ determines the observed spatial scale. In order to include the interaction between scales, the CML assumes that there is a fully developed turbulence state, and this state is characterized by the existence of eddies in different spatial scales. Here, we labeled these scales by an index $k$. For $k=1$, the model contemplates eddies in the largest spatial scale, $\ell_0/2$, which are moving through a medium surrounded by other smaller eddies with size $\ell_0/2^{k}$ in the inertial range. We model the first scale by the Langevin force equation in \eqref{langevin}, with a chaotic forcing given by
\begin{equation}
    \zeta(t) = (\gamma\tau)^{1/2}\sum_{n = 1}^{\infty} x_n\,\delta\left( t - n\tau \right) \ .
    \label{force}
\end{equation}
Here, we interpret the characteristic time $\tau$ as the time average between ``collisions" or interactions in the largest scale, which could be produced by an arbitrary driving force in the system, and $x_n$ is the normalized amplitude of the fluctuations or noise. In order to study the formation of heavy tailed and asymmetric VDFs, we consider that $x_n$ follows the chaotic Ulam map, $x_{n+1} = 1 - 2x_n^2$, which keep the fluctuations in the range of -1 to 1, with a probability distribution given by $f(\zeta) = \left[\pi^2(1 - \zeta^2)\right]^{-1/2}$. As shown by Beck~\citet{beck1994, beck1996dynamical}, the use of the chaotic Ulam map is necessary to model the chaotic forces that produces skewed VDFs in the flow.

Since the smallest timescale is $\tau$, it is convenient to discretize equation \eqref{langevin} using \eqref{force}. Therefore, integrating the Langevin equation in a small time interval $\Delta t = \tau$, we obtain
\begin{equation}
\label{cml1}  
     v_{n+1} = \lambda v_{n} + \sqrt{\gamma\tau}\, x_n \ ,
\end{equation}
where we define $v_n$ as $v(\tau n)$ and $\lambda = e^{-\gamma\tau}$. Further, to model the interactions between scales we consider the conservation of momentum between scales $k-1$ and $k$. The basic assumption is that the momentum loss at level $k-1$ serves as a driving force at level $k$. Moreover, daughter eddies at level $k$ gets only a random fraction $\xi_n^{k-1}$ of the momentum loss of the mother eddy at level $k-1$, and other part is dissipated. Thus, the momentum balance for the eddies at scale $k$ is given by
\begin{equation}
\label{cmlk} 
    m_k v_{n+1}^{(k)} = \lambda_k m_k v_{n}^{(k)} + \xi_n^{(k-1)} m_{k-1}\left( 1 - \lambda_{k-1} \right) v_{n}^{(k-1)}  \ ,
\end{equation}
where $m_k$ and $m_{k-1}$ are the masses of eddies at scales $k$ and $k-1$, respectively, $\lambda_k = e^{-k\gamma\tau}$ is a constant damping at level $k$. Clearly, here we approach the breaking factor linearly in $k$. In addition, we considered $\xi_n^{(k-1)}$ as a uniform random distribution with values between 0 and 1. This parameter can be understood as a random fraction of transferred momentum from the scale $k-1$ to the next scale $k$. In other words, the term $m_{k-1}(1-\lambda_{k-1})v_n^{(k-1)}$ represent the lost of momentum by friction in the $k-1$ scale that serves as a driving force for the next scale, $k$. Moreover, as the friction at the scale $k$ is produced by several smaller eddies, only a fraction of energy is transferred to one eddy in the next scale. It is important to mention that other choices for $\xi_n^{(k-1)}$ may be considered. However, as argued by \citet{beck1994}, and numerically shown by \citet{gallo2022langevin}, those details do not significantly affect the properties of the VDFs in the stationary state, which is the focus of our study.

Then, combining Eqs.~\eqref{cml1} and \eqref{cmlk}, in discrete form, we obtain the Coupled Map Lattice model. The equations are given by
\begin{eqnarray}
    \label{CML0}
    & v_{n+1}^{(1)} = \lambda_1 v_{n}^{(1)} + \sqrt{\gamma\tau}\, x_n \ , \\
    & v_{n+1}^{(k)} = \lambda_kv_{n}^{(k)} + c_k\ \xi_n^{(k-1)} \left( 1 - \lambda_{k-1} \right) v_{n}^{(k-1)} \ .
    \label{CML}
\end{eqnarray}
Here it must be noted that the factor $c_k$ represent the inverse of $\beta_k = m_k/m_{k-1}$, a characteristic parameter of the $\beta$-model~\citep{benzi1984multifractal}, which serves as a coupling factor between spatial scales, through the ratio between the masses of the eddies at the scales $k-1$ and $k$, respectively. Here, such relation comes from equation (\ref{cmlk}). In addition, the value of $c_k$ is also related to the width of the PDF at the scale $k$, which is related with the thermal speed of the system in the case of velocity fluctuations. As mentioned by \citet{beck1994}, the relation between the $c_k$ values and formation and properties of heavy tails in the PDF is weak. Therefore, for the sake of simplicity and to focus on the description of the tails of the PDFs, it is customary to select $c_k = c$, constant for all scales. Hence, in order to compute in a simulation, we consider $c=2$ for all scales. This means that, on average, a mother eddy preferentially interacts with eddies what are half its mass. Moreover, $v^{(k)}_n$ can be interpreted as velocity of the center of mass of an eddy in the $k$-scale at time $n\tau$. Finally, we notice that the system is determined by two characteristic times, which are condensed in the parameter $\gamma\tau$. This factor is actually the ratio between these characteristic time scales. First we have $\tau$, the time scale of fluctuations in the large scale $\sim \ell_0$, given by $\tau \sim \ell_0/w$, where $w$ is the thermal velocity of the fluctuations. On the other hand, in the Langevin Equation (\ref{langevin}), $\gamma$ is the dynamic viscosity given by $\gamma \sim \nu_0/\ell_0^2$, with $\nu_0$ the kinematic viscosity. Then, we have $\gamma\tau \sim \nu_0/w\ell_0 \sim $ $R_\lambda^{-1}$. Therefore, $\gamma\tau$ factor is inversely proportional to the Reynolds number of the fluid, $R_\lambda$~\citep{beck1994,beck1996dynamical}. Then, as $\gamma \tau$ is the only free parameter of the model, in the CML, the dynamics of the system is determined by the level of turbulence of the media, measured by $R_\lambda$. The CML equations~\eqref{CML} represent a straightforward model to study the consequences of turbulence as a one dimensional system at different scales.
\begin{figure}[htb]
\centering
    \includegraphics[width=0.8\textwidth]{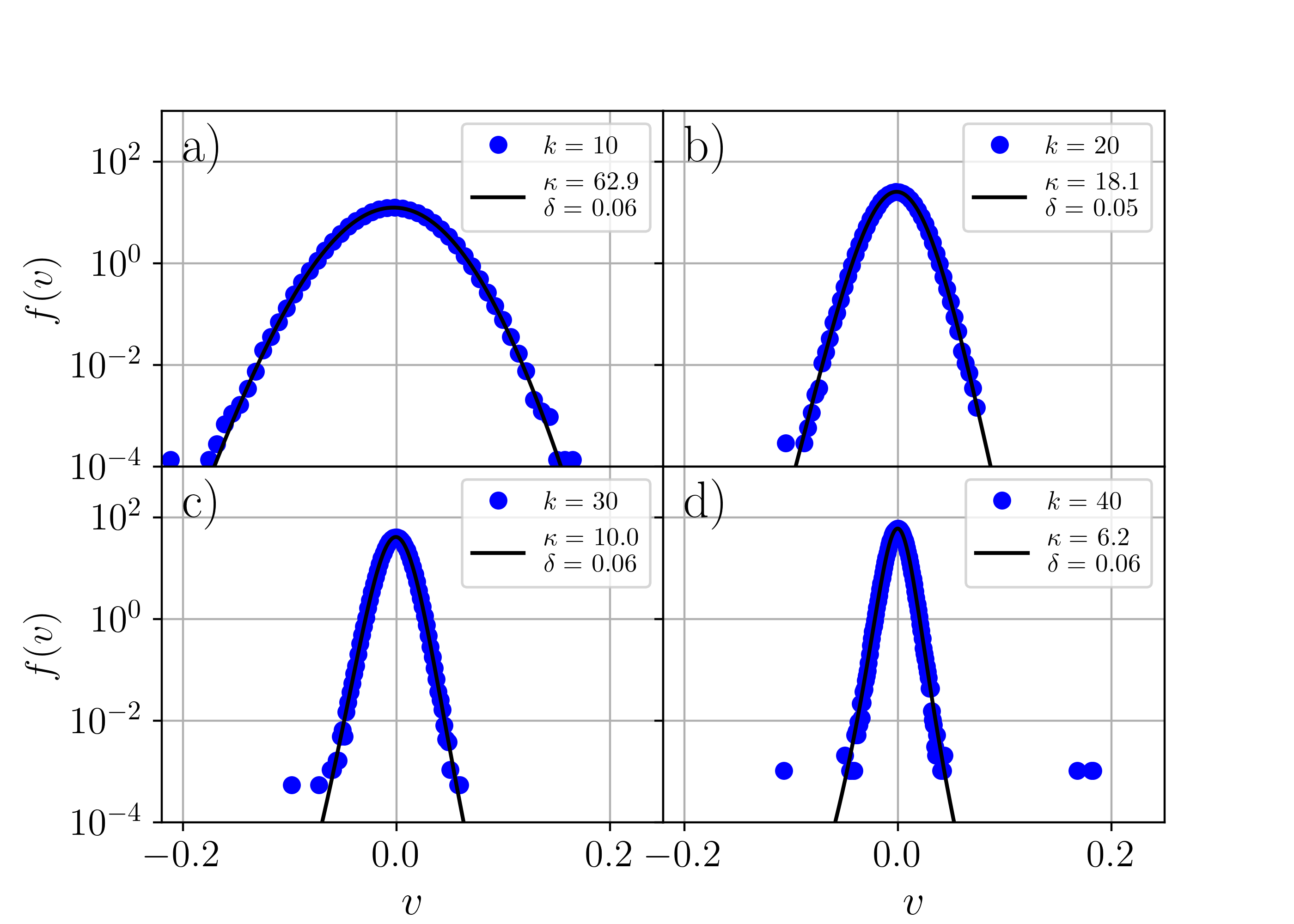}
    \caption{VDF in the steady state of the CML system in Eq. \eqref{CML}, at different $k$-scales and $R_\lambda = 10^{3.0}$. The blue dots represent the VDF of the data obtained by the simulations, and the solid black lines is the best fit using a Skew-Kappa distribution in Eq.\eqref{f_fit}. In each panel the scale $k$, the $\kappa$ and $\delta$ parameter are: (a) $k = 10$, $\kappa = 62.9$, $\delta = 0.06$, (b) $k = 20$, $\kappa = 18.1$, $\delta = 0.05$, (c) $k = 30$, $\kappa = 10.0$, $\delta = 0.06$, and (d) $k = 40$, $\kappa = 6.2$, $\delta = 0.06$.  }
    \label{VDF}
\end{figure}

\section{Analysis and Results}
\label{sec:AyR}
To numerically solve the CML equations \eqref{CML} we consider an ensemble of $N = 10^6$ eddies in each of the $k$ scale, from $k = 1$ up to $k_\text{max} = 50$, taking into account that the coupling factor between scales is constant through all scales ($c_k = 2$). In addition, to test our model with different levels of turbulence (measured by the Reynolds number) we solve the equations using five different values for $\gamma\tau$, between $10^{-4}$ and $10^{-2}$, or, similarly, using different values of Reynolds number, between $10^{2}$ and $10^{4}$.

As initial condition we select a delta function distribution for velocities centered at $v^{(k)}_0=0$, and evolve the CML equations in time until the system reaches a steady state with constant energy $E$. Namely,
\begin{align}
    \frac{dE}{dt} = 0, \quad \text{where} \quad E = \sum_{k = 1}^{k_\text{max}} \frac{1}{2}m_k \langle v^2 \rangle_k \ .
    \label{eq}
\end{align}

On the other hand, following the observations of heavy tails in the PDF of fluctuations in space plasmas \citep{chasapis2018situ, chhiber2020clustering}, fitting them with Kappa distributions by \citet{pollock2018magnetospheric}, and the study of the skewness in the distribution \citep{beck2000,rizzo2004environmental,zenteno2021model,zenteno2022role}, we fit the VDF in all $k$ scales with a Skew-Kappa distribution, as in Eq. (\ref{f_fit}), where $A_{\kappa,\delta}$, $w$, $\kappa$ and $\delta$ are fit parameters. We had special attention to the relation between these quantities respect to the scale $k$ for different values of $R_\lambda$. Skew-Kappa distributions correspond to a generalization of the Maxwellian distribution mainly in two ways. The first is that it is widely used to model out-of-equilibrium systems in which the VDF exhibits large kurtosis due to powerlaw supra-thermal tails for higher energies (or velocity). Under such context, Eq.~\eqref{f_fit}, depending on the value of the $\kappa$ parameter the VDF will model the distribution with a quasi-thermal core (for $v\le\sqrt{\kappa}w$), and power-law tails for larger velocities. The extent and energy of the tails increases with decreasing $\kappa$, and the VDF collapses to a Maxwellian in the limit $\kappa\to\infty$. And the second, Skew-Kappa distributions, as its name says, have a skewness parameter, $\delta$, which is related to the turbulent nature of the system \citep{beck2000,zenteno2021model}. Furthermore, it should be noted that this fit is only valid for small values of skewness parameter, i.e. $\delta \ll 1$, and in a limited range of $v$ such that $v/w \le |2/\delta+\delta/8|$. 

Figure \ref{VDF} shows the VDF of steady-state solution of the CML system at $R_\lambda = 10^{3.0}$ and different $k$-scales at $k=$ 10, 20, 30 and 40. Blue dots represent the VDF obtained with the simulations, and solid black lines the best fit using Eq. \eqref{f_fit}. From the figure we note a good fit by the function used (the Skew-Kappa VDF). In addition, it is clear that $\kappa$ decreases with increasing $k$, such that $\kappa=62.9\pm 0.1$ for $k=10$, and $\kappa=6.2 \pm 0.1$ for $k=40$, in the same way as in \citet{gallo2022langevin}. On the other hand, we can observe that the skewness parameter, $\delta$, keeps its value approximately constant; in this case at $\delta = 0.06$.

\begin{figure}[htb]
\centering
    \includegraphics[width=0.8\textwidth]{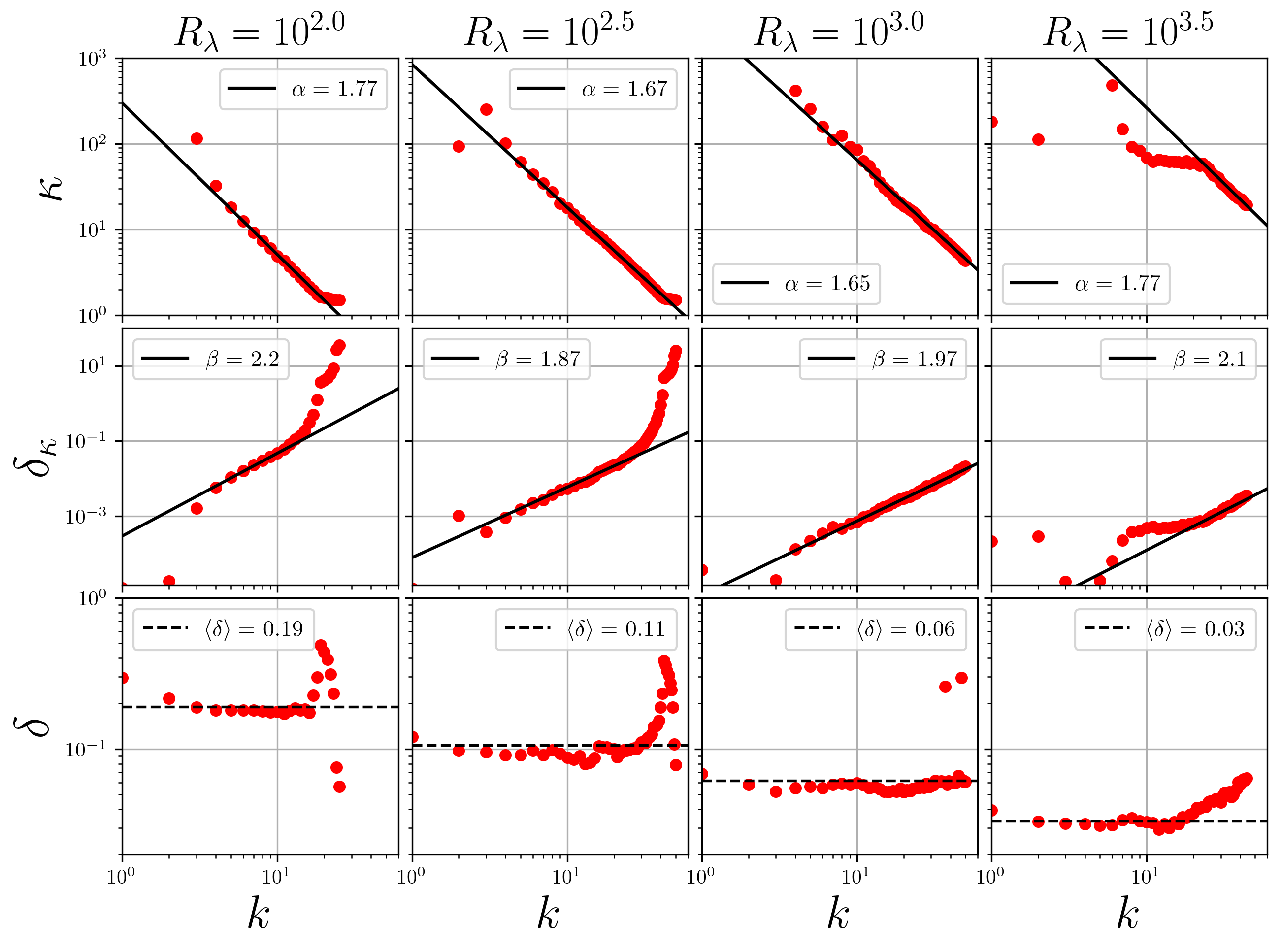}
    \caption{Comparative plot of the values of the fitting parameters such as $\kappa$ (top), $\delta_\kappa$ (middle) and $\delta$ (bottom), depending on the spatial scale $k$, for each value of Reynolds number, $R_\lambda = 10^{2.0}, 10^{2.5}, 10^{3.0}$ and $10^{3.5}$. For the first two variables a power law function with exponents $-\alpha$ and $\beta$ is overploted to the data data, respectively. The last variable, $\delta$, is compared with its average value in each case.}
    \label{panel}
\end{figure}

In order to characterize this behaviour, we performed a systematic analysis of the relation between $\delta$, $\kappa$, $R_\lambda$ and the scale $k$. Top panels of Figure \ref{panel} show the adjusted $\kappa$ as a function of $k$ for different values of $R_\lambda = (\gamma \tau)^{-1}$ (from left to right, $R_\lambda=$ 10$^{2.0}$, 10$^{2.5}$, 10$^{3.0}$ and 10$^{3.5}$, respectively). It should be noted that with this selection of Reynolds number values we were simulating the behavior of a plasma system in the context of space plasmas, as the solar wind, the plasma sheet, among others. Due to numerical issues, we could not evaluate the dynamics of Reynolds numbers above $10^4$. This limitation was due to the numerical resolution required to accurately simulate the behavior of the plasma system. However, regarding the range of Reynolds numbers that we were able to consider, we can find that the effective Reynolds number for the solar wind was determined to be $260.000 \pm 20.000$ at 1 AU from the Sun as shown by \cite{weygand2007taylor}, but smaller (less than 10$^4$) specially at larger solar distance as reported in \cite{Parashar2019}. Similarly in the case of the the plasma sheet inside the magnetosphere, \cite{weygand2007taylor} found Reynolds number values between $7 \pm 1$ and $110 \pm 12$, also in line with our sampling of values.

From the figure \ref{panel} we can see that, in general, for all values of $R_\lambda$, $\kappa$ exhibits large values for small $k$, or, in other words, greater spatial scales. This can be explained because we should expect a Brownian-like motion for the largest scale ($k=1$), represented by a Maxwellian VDF in the steady state, i.e. $\kappa \rightarrow \infty$. Therefore, it would be difficult to obtain an accurate value for $\kappa$. On the other hand, as we move on the $k$-scale, towards high values, $\kappa$ rapidly decreases, up to the saturation scale (largest $k$) in which $\kappa \to 3/2$ (the smallest possible $\kappa$ value in our model). Therefore, as $k$ increases, the $\kappa$ parameter is finite and smaller. That is, as the scale size decreases the system moves away from a Maxwellian regime and the fluctuations distribution is well represented by a distribution with heavy tails, such as Skew-Kappa distributions.

In addition, regarding $\kappa$ parameter, we observe a clear decaying power law behavior $\kappa \sim k^{-\alpha}$. For each value of $R_\lambda$ the best power law is also included as a superposed black line in each panel. Remarkably, for all considered values of Reynolds number, $\alpha\approx 1.74$. Furthermore, from the figure it is also possible to notice how this power law behaviour also depends on the value of $R_\lambda$, as, in general, $\kappa$ increases with increasing the Reynolds number. However, from Figure~\ref{panel} we can recognize two regimes where the purposed power law model is no longer representative: first, for low values of $k$ (or high spatial scales), given that the points are expected to be dispersed since the system tends to the Maxwellian state. As Maxwellian is the limit when $\kappa\to\infty$, when the system is more Maxwellian than Kappa distributed, the algorithm found high values for $\kappa$ (larger than $10^2$), and therefore there is low precision in the convergence of the fit method. Second, for high values of $k$ (small spatial scales), where the points saturate to a fixed value of 3/2. Since the $\kappa$-like fit we use has a singularity at this value, where the tails of the PDF seem too large to be described using a Kappa distributions. In such cases, other models may provide a better description of the PDFs. 

It is important to mention that the validity range in $k$ also depends on the Reynolds number, as shown on the top of the panel. In short, the slope seem to be constant for all values of $R_\lambda$, but we can see how the curve moves to the right for increasing Reynolds number, and also the power law will be presented as long as $3/2<\kappa \lesssim 10^2$. For all other values of $\kappa$, the trend is broken. We attribute this behavior of the Kappa parameter to the correlation gain as we advance in the $k$ index. Let us recall that scale $k$ contains part of the information of all $k-1$ previous ones. In this sense, we can say that smaller scales are more correlated. Therefore, the higher the level of correlation, the smaller the value of $\kappa$. This is consistent with the arguments given by \citet{livadiotis2021thermodynamic}. There, the authors discuss the relationship between $\kappa$ and the level of correlation of a system. They justify a non-additive entropy since a correlation term in the total entropy which depends on the entropy of two subsystems and $\kappa$. A turbulent system, where actually exist high spatio-temporal correlations, could be  easily explained by this point of view. Then, it is feasible to measure fluctuation distributions in turbulent systems, with heavy tails or large kurtosis, which could be well fitted with Kappa-like distributions~\citep{gallo2022langevin}. 

In order to include more information about our system, we defined $\delta_\kappa = \delta/(\kappa - 3/2)$, which is indeed a combination of two fit parameters, such as $\delta$ and $\kappa$. This definition is not a coincidence since $\delta_\kappa$ is naturally generated from Eq. \eqref{f_fit}. In this way, we can interpret $\delta_\kappa$ as an effective skewness parameter. As we can notice, on the bottom of Figure \ref{panel}, we plot $\delta$ as a function of the scale level, $k$. As expected, due to the arguments exposed by \citet{beck2000}, we should not see a dependence between skewness parameter and the scale, but only on the Reynolds number, namely, $\delta \sim R_\lambda^{-1/2}$. In fact, we can see, from the plot of $\delta$ parameter, a non-strong-dependence on the scale. Therefore, it is expected that, if $\kappa$ decays as a power law, for large $k$, then $\delta_\kappa$ will be a power law with a positive slope, just as we see in the Figure mentioned.

\begin{figure}[htb]
\centering
    \includegraphics[width=0.8\textwidth]{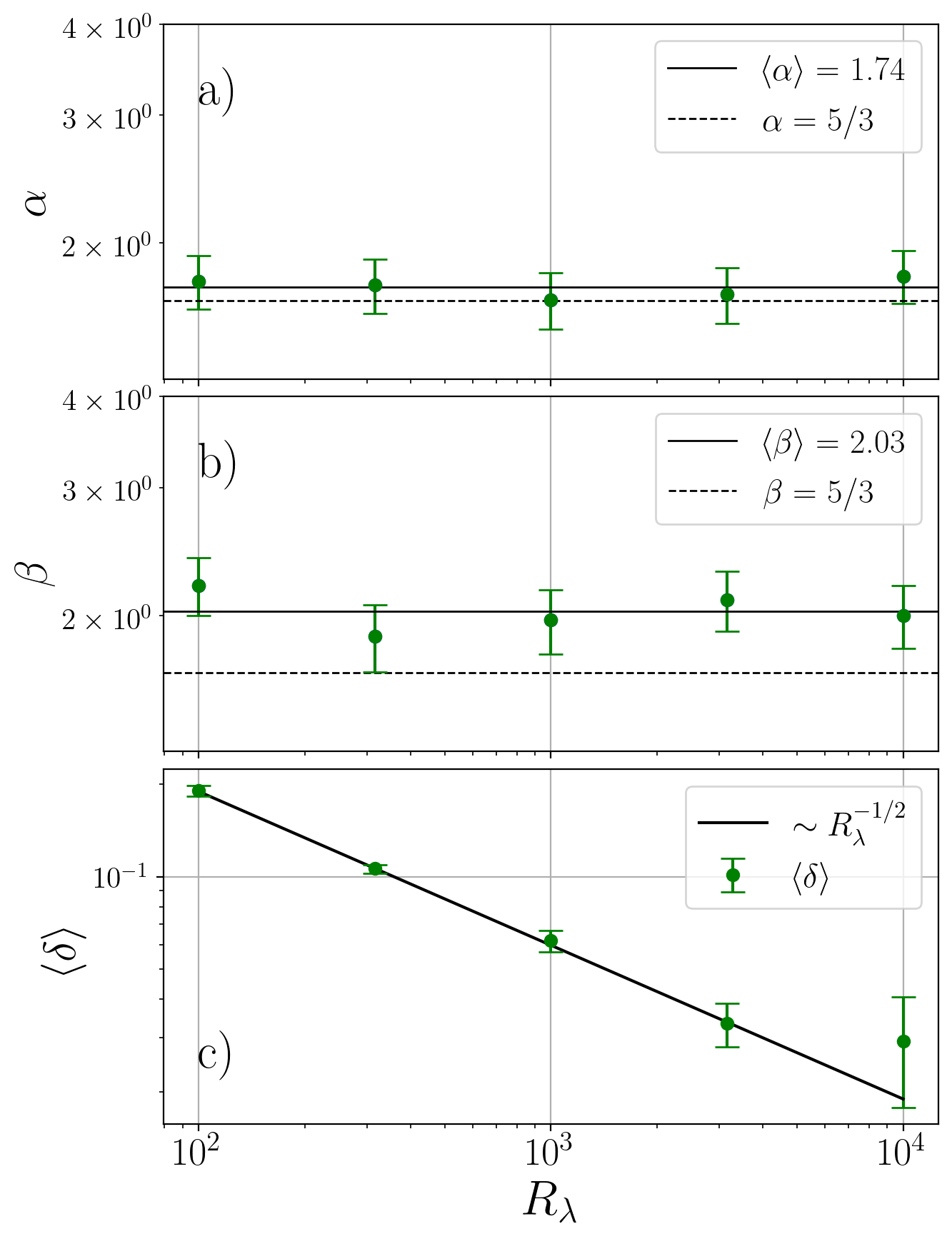}
    \caption{Obtained values of slopes in the power law fit for a) Kappa-parameter $\kappa$, b) effective skewness parameter $\delta_\kappa$. And finally, on the bottom, c) the average value for $\delta$ for each value of Reynolds number, $R_\lambda$. At the top, a), and in the middle, b),  we compare the average value of the data obtained from the fits with $5/3$, as a reference value. At the bottom, in c), we fit the data obtained from the average values of $\delta$ with a power law. The slope for this fit gave us a value of -1/2, as the theory suggest.}
    \label{slopes}
\end{figure}
As the shape of the data suggest, we fit a power law for the effective skewness, namely, $\delta_\kappa \sim k^{\beta}$. Due to $\delta_\kappa$ decays as the inverse of $\kappa$, and $\delta$ is apparently constant in a certain range, we should expect $\alpha \approx \beta$ in that range of $k$. Although with some error this is fulfilled. As we can see in the figure, $\delta_\kappa$ is well fitted by a power law in the same range as $\kappa$, but there is a small discrepancy between the mean value of $\alpha$ and $\beta$, given by $\langle \beta \rangle - \langle \alpha \rangle = 0.29$. Nevertheless, this difference is easily attributable to the sensitivity of the $\delta$ parameter when we measure it. Indeed, we can see the above in more detail in Figure \ref{slopes}a and \ref{slopes}b, where we plot the values of slopes $\alpha$ and $\beta$ for each value of $R_\lambda$, and we compare them with our reference value 5/3, which is apparently a constant value as studied in \citet{gallo2022langevin}. Finally, in Figure \ref{slopes}c, we plot the values of skewness parameter, $\delta$, for each value of Reynolds number, $R_\lambda$. In order to compare these values with the theory, we overplot the curve $R_\lambda^{-1/2}$. We notice that there is a correspondence between the data and the theoretical curve, except for the last point at $R_\lambda = 10^{4.0}$, which is likely to have primarily numerical error. Since in that case the $\gamma \tau$ number in the equations \eqref{CML} is quite small, it is expected to generate numerical issues in the convergence of the system.  

Furthermore, considering the relation between the scale $k$ and the $\kappa$-parameter, our numerical results indicate that
\begin{equation}
    \kappa(k) \sim   k^{-5/3} \ ,
    \label{final}
\end{equation}
is the relation between $\kappa$ and the spatial scale. Therefore, the equation (\ref{final}) is a functional relation which is valid for a certain range of spatial scales, according to the CML. And not only that, since as we see on the bottom of the Figure \ref{slopes}, we conclude that, indeed, the skewness parameter relates with the level of turbulence, namely, 
\begin{equation}
    \delta(R_\lambda) \sim \ R_\lambda ^{-1/2} \ .
    \label{delta}
\end{equation}
This last result could be a very useful tool at the time to measure the level of turbulence in a certain system. As the same way in  \citet{beck2000, rizzo2004environmental}, when we have two measuring tools ($i = \,$\{$1$, $2$\}), at two different points, separated by a characteristic length $\ell$, which measure a variable $u_i$, we could construct the difference $v = u_1 - u_2$ (as in Eq. \eqref{diffV}), framed within a PVI quantity. In this way we can obtain the PDF of PVI, such as the authors show in their works. There, it is clear that when the distribution is built, it has a slight asymmetry, which can well fitted by a Skew-Kappa distributions. The latter allows an indirect measurement of the level of turbulence through the quantification of the skewness parameter $\delta$. 

\section{Conclusions}
\label{sec:sum}

In this study we have presented a numerical work about the velocity fluctuation distribution generated in turbulent flows at different scales $k$ and different levels of turbulence, $R_\lambda$, both quantities defined in section \ref{sec:model}. For that purpose, we used a Couple Map Lattice model (CML) which considers turbulence as a coupled Langevin system at different scales with a chaotic driving force modeled using the Ulam Map. With such mode we found that the PDF of velocity fluctuations at steady state in different scales can be well represented by Skew-Kappa distribution. With this fit, we could relate that $\kappa$ depends on the scale $k$ and the level of turbulence represented by the Reynolds number (as reported in \citet{gallo2022langevin}). On the other hand we observe that the skewness, quantified by $\delta$ parameter in a Skew-Kappa distribution, is also strongly related with the level of turbulence but not with the scale $k$. 

The numerical results of our study show the relation between the spatial scale, the Reynolds number, and the $\kappa$ and $\delta$ parameters. We observe how $\kappa$ exhibits a power law behavior given by Eq. \eqref{final}. In that regard, we can conclude that the nature of the force used in the model does not change the shape of the dependence between $\kappa$ and the spatial scale. In this way, when the Reynolds number grows, the $\kappa$ parameter also increases. This fact is expected since in a system with large enough Reynolds number, strong turbulence should inhibit the internal correlations leading to non Maxwellian PDFs.  Secondly, we compare the values of skewness, parameterized by $\delta$, with respect to the spatial scale $k$. It is possible to notice that the asymmetry is weakly dependent on the scale, in the same way as \citet{beck2000} suggest and is concluded in \citet{gallo2022langevin}. This is because the theory suppose that $\delta$ does not depend on the $\kappa$ parameter i.e. not depends on the scale either. Therefore we are able to found a match between the simulation approach with theoretical works.  

While our research primarily focuses on other aspects of the system dynamics, it is also worth mentioning that our model exhibits the characteristics of the inertial range. Specifically, our model reaches equilibrium at larger spatial scales initially and gradually achieves a stationary behavior at progressively smaller scales, with clear scaling laws between $\delta_\kappa$, $\kappa$, and $k$. However, as all considered scales follow the same equations, the model is not able to exhibit a spectral brake between the inertial and kinetic ranges as observed in the solar wind. To do so we should have to introduce new rules to Equations (\ref{CML0}) and (\ref{CML}). Moreover, other characteristic of turbulent flows (as intermittency) are also not considered. We acknowledge that this may happen but this it is not the primary focus of our work, and we have not thoroughly explored these phenomena in our article. In summary all these issues are relevant and worth to study, but investigating the potential intermittent behavior of plasma fluctuations, the properties of the inertial range, and the possible inclusion of a spectral break, are all beyond the scope of our current study.

That being said, as far as $\kappa$ parameter is concerned, the results are qualitatively consistent with the observational study by \citet{pollock2018magnetospheric}, where the Partial Variance of Increments of the velocity and magnetic field fluctuations were calculated in the context of space plasmas. In addition in the case of neutral fluids, we have the data obtained and analyzed by \citet{rizzo2004environmental}, where the authors did the study with temporal and spatial PVI. Respect to the temporal PVI, the conclusions are clearly related with the findings presented in \citet{gallo2022langevin}. It is possible to notice the qualitative relation between our results about $\kappa$ and the characteristic time lag ($\tau$) that they used in their work. If the time lag increases then $\kappa$ also increase its value. On the other hand, for the other definition of PVI, which is construct with a characteristic length scale, the results seem to be quite similar to those reported by \citet{beck2000}, who analyzed data from a jet experiment of turbulent flows. 

To complement those studies, we propose the systematization of the quantification of skewness parameter in PDFs of spatial PVI, evaluating the fit for several fluids with different Reynolds numbers. Not only that, but also we suggest that the methodology exposed in this article could be included in future observational works in order to extend the study of PVI in space plasma. And though this has been explored in studies as \citet{chasapis2018situ}, we emphasize the importance of making the evaluation on the PDF without absolute value by making the difference between both measurements, and thus obtain an asymmetric distribution.  
In summary, on the basis of a simple numerical model of a turbulent system we have found that the PDFs of velocity differences fluctuations follow Skew-Kappa distributions, and that there is a robust relation between the scaling of $\kappa$, the skewness parameter,$\delta$, the spatial scale level, $k$, and the Reynolds number of the flow. We expect these results to be useful to characterize turbulent systems in different contexts, specially in space plasma environments, where the Reynolds number is not always easy to obtain~\citep{Parashar2019}, and our numerical predictions to be tested by spacecraft observations or in experimental setups.

\section*{Acknowledgements}

We are grateful for the support of ANID, Chile through a National Doctoral Scholarship No. 21182002 (IGM), and FONDECYT grant No. 1191351 (PSM).  





\bibliography{preprint}{}
\bibliographystyle{aasjournal}



\end{document}